\documentstyle[12pt]{article}

\newcommand{\be}{\begin{equation}}
\newcommand{\ee}{\end{equation}}
\newcommand{\ben}{\begin{eqnarray}}
\newcommand{\een}{\end{eqnarray}}

\def\simgt{\rlap{\lower 3.5 pt\hbox{$\mathchar \sim$}}\raise 1pt \hbox {$>$}}
\def\simlt{\rlap{\lower 3.5 pt\hbox{$\mathchar \sim$}}\raise 1pt \hbox {$<$}}

\begin{document}
\baselineskip=24pt
%
\title{
{\Large Lattice QCD Calculation of the Kaon $B$-parameter \\ 
with the Wilson Quark Action\vspace*{0.5cm}}}

\author{JLQCD Collaboration \vspace*{0.2cm}\\ 
        S.~Aoki$^{\rm a}$, M.~Fukugita$^{\rm b}$, 
        S.~Hashimoto$^{\rm c}$, N.~Ishizuka$^{\rm a}$,
        Y.~Iwasaki$^{\rm a,d}$, \\ K.~Kanaya$^{\rm a,d}$, 
        Y.~Kuramashi$^{\rm e}$, M.~Okawa$^{\rm e}$, 
        A.~Ukawa$^{\rm a}$, T.~Yoshi\'{e}$^{\rm a,d}$ 
\vspace{5mm}\\ 
  {\it  Institute of Physics,  University of Tsukuba$^{\:a)}$, }\\
  {\it  Tsukuba, Ibaraki 305, Japan }
\vspace{3mm}\\
  {\it  Institute for Cosmic Ray Research, University of Tokyo$^{\:b)}$, }\\
  {\it  Tanashi, Tokyo 188, Japan }
\vspace{3mm}\\
  {\it  Computing Research Center$^{\:c)}$, }\\
  {\it  High Energy Accelerator Research Organization(KEK), }\\
  {\it  Tsukuba, Ibaraki 305, Japan } 
\vspace{3mm}\\
  {\it  Center for Computational Physics,  University of Tsukuba$^{\:d)}$, }\\
  {\it  Tsukuba, Ibaraki 305, Japan } 
\vspace{3mm}\\
  {\it  Institute of Particle and Nuclear Studies$^{\:e)}$, }\\
  {\it  High Energy Accelerator Research Organization(KEK), }\\
  {\it  Tsukuba, Ibaraki 305, Japan } 
}

\date{}
\maketitle
 
\begin{abstract}
\baselineskip=24pt

The kaon $B$ parameter is calculated in quenched lattice QCD
with the Wilson quark action.  The mixing problem
of the $\Delta s=2$ four-quark operators is solved non-perturbatively 
with full use of chiral Ward identities, and this method enables us to
construct the weak four-quark operators exhibiting good chiral behavior.
We find $B_K({\rm NDR}, 2{\rm GeV})=0.562(64)$ in the continuum limit,
which agrees with the value obtained with the Kogut-Susskind 
quark action.  

\end{abstract}

\newpage

Reliable knowledge of the $K^0-\bar K^0$ transition matrix element
$B_K$ is an indispensable ingredient for further advancement in
$CP$ violation phenomenology, and the evaluation of this $B_K$ 
parameter has been one of the major targets for
lattice QCD calculations. Indeed, lots of effort has  
been expended 
towards this end. The successful calculations of $B_K$ so far
achieved \cite{bk_ks,saoki} exclusively employ the Kogut-Susskind quark action
that respects chiral $U(1)$ symmetry. Whereas the verification whether both 
Wilson and Kogut-Susskind quark actions yield the identical result for $B_K$
is an important step
to give a full credit to the lattice QCD calculation, 
the attempts made with the
Wilson quark action have not yielded much
success\cite{latt88,bk_w,romeBK}:
the use of the Wilson action with explicit breaking of chiral symmetry 
causes mixing among four-quark operators of different chiral 
structure, 
and ensuring the correct chiral behavior of the $\Delta s=2$ operators 
is substantially more complicated than with the Kogut-Susskind action. 
Early studies have shown that the mixing problem is not adequately
treated by perturbation theory, leading to an ``incorrect answer''
for the matrix element \cite{latt88}. Attempts were then made to
solve the mixing problem non-perturbatively with the aid of 
chiral perturbation theory\cite{bk_w}.  Unfortunately, they were not 
successful since the calculation contains large systematic uncertainties 
arising from higher order effects that survive even in the continuum limit.
More recently a proposal is made \cite{romeBK} to improve
the chiral behavior of the $\Delta s=2$
operator with the use of non-perturbative 
renormalization (NPR) \cite{npr}.
The mixing coefficients derived in this approach, however, depend
sharply on the momentum scale where the operator is to be
evaluated, and this causes a subtlety in estimating the matrix element. 

In this Letter we propose a non-perturbative method to solve the 
operator mixing problem with use of chiral Ward identities\cite{wi}.
This method fully incorporates the chiral properties 
of the Wilson action, yielding the $\Delta s=2$
operator that shows good chiral behavior. No effective theories
are invoked to estimate the matrix element.
The final result we obtained is encouraging: it shows good
agreement with the result with 
the Kogut-Susskind quark action.  
We shall also revisit the perturbative method.

Let us consider a set of weak operators in the continuum  
$\{\hat O_i\}$ which closes under chiral rotation
$\delta^a \hat O_i=ic^a_{ij}\hat O_j$.  These operators are
given by linear combinations of a set of lattice operators
$\{O_\alpha\}$, as $\hat O_i=\sum_\alpha Z_{i\alpha}O_\alpha$.
We choose the mixing coefficients $Z_{i\alpha}$ such that
the Green functions of $\{\hat O_i\}$ with quarks in
the external states satisfy the chiral
Ward identity to $O(a)$.  This identity can be derived in a
standard manner\cite{wi} and takes the form
\ben
-2\rho Z_A\langle\sum_xP^a(x)\hat O_i(0)
\prod_k\tilde\psi(p_k)\rangle  
+c^a_{ij}\langle\hat O_j(0)\prod_k\tilde\psi(p_k)\rangle 
\label{eq:wi}\\
-i\sum_l\langle\hat O_i(0)\prod_{k\ne
l}\tilde\psi(p_k)\delta^a\tilde\psi(p_l)\rangle+O(a)=0, \nonumber
\een
where $p_k$ is the momentum of the external quark,  $Z_A$
and $\rho=(m-\delta m)/Z_A$ are constants to be determined from the 
Ward identities for the axial vector currents\cite{rhoza},
and $P^a$ is the pseudoscalar density of flavor $a$.
We note that the first term comes from the chiral variation 
of the Wilson quark action and the third represents the chiral 
rotation of the external fields.

The four-quark operator relevant for $B_K$ is given by $\hat O_{VV+AA}=VV+AA$  
where $V=\bar s\gamma_\mu d$ and $A=\bar s\gamma_\mu\gamma_5 d$. 
Then, $\hat O_{VV+AA}=VV+AA$ and  $\hat O_{VA}=VA$ form a minimal set of 
the operators that closes under $\lambda^3={\rm diag}(1,-1,0)$ chiral rotation.
Taking account of $CPS$ symmetry(note that we take $m_d=m_s$ in
this article)\cite{latt88},
mixing of these operators is written
${\hat O_{VV+AA}}/2$=$Z_{VV+AA}\left(O_0+z_1O_1+ \cdots
+z_4O_4\right)$ and ${\hat O_{VA}}=Z_{VA}\cdot z_5O_5$, where the six lattice 
operators $O_i$ are given by 
$O_0 = \left(VV + AA\right)/2$,
$O_1 = \left(SS+TT+PP\right)/2$,
$O_2 = \left(SS-TT/3+PP\right)/2$,
$O_3 = \left(VV-AA\right)/2 + \left(SS-PP\right)$,
$O_4 = \left(VV-AA\right)/2 - \left(SS-PP\right)$ 
and
$O_5=VA$ with $S=\bar s d$, $P=\bar s\gamma_5 d$ and 
$T=\bar s[\gamma_\mu, \gamma_\nu] d/2$ $(\mu < \nu)$.   
These operators are in the Fierz eigenbasis which we find convenient 
when taking fermion contractions for evaluating the Green functions in (1). 

We consider the four external quarks having an equal momentum
$p$, and denote by $\Gamma_{VV+AA}$ and $\Gamma_{VA}$ the sum of the
Green functions on the left hand side of (\ref{eq:wi}) with external
quark legs amputated. 
Using the projection operator $P_i$ 
for the Fierz eigenbasis corresponding 
to $O_i$, we write $\Gamma_{VV+AA}/Z_{VV+AA}=\Gamma_5P_5$ 
and $\Gamma_{VA}/Z_{VA}=\Gamma_0P_0+\Gamma_1P_1+\cdots
+\Gamma_4P_4$. 
Writing ${\hat O_{VV+AA,VA}}$ in  (\ref{eq:wi})
in terms of lattice operators,
we obtain six equations for the five coefficients $z_1,\cdots,z_5$:
\be
\Gamma_i=c^i_0+c^i_1z_1+\cdots +c^i_5z_5=O(a), \quad i=0,\cdots, 5
\ee 
This gives an overconstrained set of equations, and 
we may choose any five equations to exactly vanish to solve 
for $z_i$: the
remaining equation should automatically be satisfied to
$O(a)$. We choose four equations to be those for 
$i=1,\cdots,4$, since $O_1,\cdots,O_4$ do not
appear in the continuum.
The choice of the fifth equation, $i=0$ or 5,
is more arbitrary. We have confirmed that either $\Gamma_0=0$
or $\Gamma_5=0$ leads to a
consistent result to $O(a)$ for $z_1,\cdots,z_4$ in the region
$pa\simlt 1$.
In the present analysis we choose $\Gamma_5=0$. 
The overall factor $Z_{VV+AA}$ is determined by the
NPR method\cite{npr}.
We convert the matrix elements on the lattice  into  those of the
$\overline{MS}$ 
scheme in the continuum using naive dimensional
regularization (NDR) renormalized at the scale
$\mu=2$GeV\cite{romeBK}:
\be
B_K({\rm NDR},\mu)=
\left(1+\frac{\alpha_s(\mu)}{4\pi}\left(-4{\rm
log}(\frac{\mu}{p})
-\frac{14}{3}+8\log 2 \right)\right)\frac{\langle
{\bar K}^0 \vert \hat O_{VV+AA} \vert K^0 \rangle}
{\frac{8}{3}\vert \langle 0 \vert {\hat A} \vert K^0 \rangle \vert^2}
\ee
where $p$ denotes the momentum at which the mixing coeffecients are evaluated.

For comparative purpose we also calculate $B_K$ with 
perturbative  mixing coefficients, for which we use the one-loop
expression in Ref.~\cite{pt} after 
applying a finite correction in conversion to 
the NDR scheme together with  
the tadpole improvement with 
$\alpha_{\overline{MS}}(1/a)$.

Let us remark here that the equations obtained in the NPR
method \cite{romeBK} corresponds to $\Gamma_i=0$ for $i=1, \cdots, 4$ 
in which the contributions of the first and the third term in the Ward 
identity (1) are dropped,  {\it i.e.,} the full chiral properties
are not taken into account in the NPR approach.
  
Our calculations are made with the Wilson quark action and the plaquette 
action at $\beta=5.9-6.5$ in quenched QCD. 
Table~\ref{tab:runpara} summarizes our run parameters. 
Gauge configurations are generated with the 5-hit pseudo heat-bath
algorithm with 2000($\beta=5.9$ and 6.1), 5000($\beta=6.3$) or
8000($\beta=6.5$) sweep intervals apart. The physical 
size of lattice is chosen to be approximately 
constant at $La\approx
2.4$fm where the lattice spacing is determined from
$m_\rho=770$MeV. At each $\beta$ 
four values of the hopping parameter are adopted such that the physical
point for the $K$ meson can be interpolated.
We estimate $m_s a/2$ from $m_K/m_\rho=0.648$ 
for degenerate $d$ and $s$ quark masses.
Errors are estimated by the single elimination 
jackknife method for all measured quantities
throughout this work.

Our calculations are carried out in two steps. We first calculate
$z_i$ and $Z_{VV+AA}$ using the quark Green functions 
having finite space-time momenta. For this purpose 
quark propagators are solved in the Landau gauge
for the point source located at the origin
with the periodic boundary condition imposed on the lattice. 
We next extract $B_K$ from the ratio
$\langle {\bar K}^0(t=T){\hat
O}_{VV+AA}(t^{\prime})K^0(t=1)\rangle / 
\langle{\bar K}^0(t=T){\hat A}(t^{\prime})\rangle
/\langle {\hat A}(t^{\prime})K^0(t=1)\rangle$,  each Green function
projected onto the zero spatial momentum,  by fitting a plateau 
seen as a function of $t^{\prime}$.
For this calculation
quark propagators are solved without gauge fixing employing the wall
source placed at the edge where the Dirichlet boundary
condition is imposed in the time
direction. We obtain $B_K$ at $m_s/2$ by quadratically 
interpolating the data at the four hopping parameters. 

We plot in Fig.~1 a typical result for the mixing coefficients 
as a function of the external quark momenta. The plot shows,
as desired, only
weak dependence of $z_i$ on momentum in the range  
$0.1 \simlt p^2a^2 \simlt 1.0$.  
This enables us to evaluate the mixing coefficients with small errors 
at the scale $p^{*}\approx 2$GeV, which
always falls within the range of a plateau 
for our runs at $\beta=5.9-6.5$.
We remark that this weak scale dependence contrasts to
the strong scale dependence obtained with  
the NPR method\cite{romeBK}.

In Fig.~2 we compare the mixing coefficients evaluated 
at the scale $p^{*}$ (filled symbols)
with the perturbative values obtained with 
$\alpha_{\overline{MS}}(1/a)$ (open symbols) as a function of lattice 
spacing. 
We remark that a large value of $z_2$ determined by the Ward identities 
sharply contrasts with the one-loop perturbative result $z_2=0$.  
For the other coefficients, the perturbative
results agree with the non-perturbative ones in sign and
rough orders of magnitude. They differ in quantitative details, however.

Let us examine the chiral property of the operator $\hat O_{VV+AA}$ by 
calculating the ratio
$\langle
{\bar K}^0 \vert \hat O_{VV+AA} \vert K^0 \rangle /
\vert \langle 0 \vert {\hat P} \vert K^0 \rangle \vert^2$, which  vanishes
at $m_q=0$ in the continuum.  
In Fig.~3 we show the results at $m_q=0$ obtained by a quadratic 
extrapolation of data in $m_q=(1/K-1/K_c)/2$, 
where WI stands for our method using chiral Ward identities and PT for  
tadpole-improved one-loop perturbation theory
(numbers are given in Table~\ref{tab:result}).
The pseudoscalar density $\hat P$ in the denominator 
is renormalized perturbatively for both cases.
A significant improvement is clearly seen with use 
of the Ward identities, the ratio becoming consistent with zero at
the lattice spacing $m_\rho a\simlt 0.3 (a\simlt 0.08$fm).
In the perturbative approach, the
chiral behavior is recovered only after extrapolation
to the continuum limit,  where we adopted a linear dependence 
on $a$ expected for the Wilson quark action in the extrapolation shown 
in Fig.~3. 
 
Our final results for  $B_K({\rm NDR},2{\rm GeV})$ are presented 
in Fig.~4 as a function of lattice spacing
(see Table 2 for numerical details). 
The method based on the Ward
identity (WI) gives a value convergent from a 
lattice spacing of $m_\rho a\approx 0.3$ ($B_K\sim0.6-0.8$). 
The large error, however, hinders us from making an extrapolation to 
the continuum limit.   Since the origin of the large error is traced to 
that of the mixing coefficients,
we develop an alternative method, which we refer to as 
WI$_{\rm VS}$, in which the denominator of the
ratio for extracting $B_K$ is estimated with the vacuum
saturation of $\hat O_{VV+AA}$ constructed by the WI method.
In this case the fluctuations in the numerator are largely
canceled by those in the denominator, and the resulting
error in $B_K$ is substantially reduced as apparent in Fig. 4. 
The cost is that the correct chiral behavior 
of the denominator is not respected  
at a finite lattice spacing due to the contributions of the pseudoscalar
matrix element besides the axial vector one.
While WI and WI$_{\rm VS}$ methods give different results at
a finite lattice spacing, the discrepancy is expected 
to vanish in the continuum limit.
A linear extrapolation in $a$ of the WI$_{\rm VS}$ results yields 
$B_K({\rm NDR},2{\rm GeV})=0.562(64)$, which we take as the best value
in the present work. This value is consistent
with a recent JLQCD result with the Kogut-Susskind
action,
$B_K({\rm NDR},2{\rm GeV})=0.587(7)(17)$\cite{saoki}.

Intriguing in Fig.~4 is that the perturbative calculation (PT),
which gives completely ``wrong value'' at $a\ne0$,
also yields the correct result for $B_K$, when extrapolated
to the continuum limit $a=0$. 
This is a long extrapolation from negative to positive, 
but the linearly extrapolated value $B_K$(NDR, 2GeV)=0.639(76) is consistent 
with those obtained with the WI or WI$_{\rm VS}$ method.
We note that a long extrapolation may bring a large error in the 
extrapolated value.

Finally we mention possible sources of systematic errors in our results 
from quenching effects and uncertainties for Gribov 
copies in the Landau gauge. 
With the Kogut-Susskind quark action it has been observed that the error
due to quenched approximation is small \cite{bk_ks,OSU}. Whether this
is supported by calculations with Wilson action we must defer to future 
studies. 
For the Gribov problem we only quote an earlier study\cite{gribov} 
which suggests that ambiguities in the choice of the Gribov copies 
induce only small uncertainties comparable to typical statistical errors 
in current numerical simulations.  
 
In conclusion our analysis for $B_K$ demonstrate the effectiveness 
of the method using the chiral Ward identities for constructing the $\Delta
s=2$ operator with the correct chiral property.  We have shown that both
Wilson and Kogut-Susskind actions give virtually the
identical answer for $B_K$ in their
continuum limit. 
We may hope that further improvement of our simulations leads to a precise
determination of $B_K$ with the Wilson quark action.  
The application of this method to $B_B$ is also straightforward.

This work is supported by the Supercomputer Project (No.1) of 
High Energy Accelerator Research Organization(KEK), and also 
in part by the Grants-in-Aid of the Ministry of Education 
(Nos. 08640349, 08640350, 08640404, 08740189, 08740221).

\newpage
\begin{center}
\section*{Tables}
\end{center}

\begin{table}[h]
\vspace{-1mm}
\begin{center}
\caption{\label{tab:runpara}Parameters of simulations.}
\vspace*{2mm}
\begin{tabular}{lllll}\hline
    $\beta$       & 5.9   & 6.1   & 6.3   & 6.5 \\ 
\hline
$L^3\times T$     & $24^3\times 64$ & $32^3\times 64$ 
                  & $40^3\times 96$ & $48^3\times 96$ \\
\#conf.           & 300             & 100             
                  & 50              & 24 \\
$K$               & 0.15862  & 0.15428  & 0.15131  & 0.14925 \\ 
                  & 0.15785  & 0.15381  & 0.15098  & 0.14901 \\ 
                  & 0.15708  & 0.15333  & 0.15066  & 0.14877 \\ 
                  & 0.15632  & 0.15287  & 0.15034  & 0.14853 \\ 
$K_c$             & 0.15986(3)   & 0.15502(2)     
                  & 0.15182(2)   & 0.14946(3) \\ 
$a^{-1}$ (GeV)    & 1.95(5)      & 2.65(11)       
                  & 3.41(20)     & 4.30(29) \\ 
$\alpha_{\overline{MS}}(1/a)$   
                  & 0.1922  & 0.1739  & 0.1596  & 0.1480 \\ 
$m_s a/2$         & 0.0294(14)   & 0.0198(16)     
                  & 0.0144(17)   & 0.0107(16) \\ 
${p^{*}}^2a^2$    & 0.9595  & 0.5012  & 0.2988  & 0.2056 \\ 
\hline
\end{tabular} 
\end{center}
\end{table}

\begin{table}[h]
\vspace{-1mm}
\begin{center}
\caption{\label{tab:result}
$\langle{\bar K}^0 \vert \hat O_{VV+AA} \vert K^0 \rangle /
\vert \langle 0 \vert {\hat P} \vert K^0 \rangle \vert^2$ 
in the chiral limit and $B_K($NDR, 2GeV) 
for WI, WI[VS] and PT methods 
as a function of $\beta$.}
\vspace*{2mm}
\begin{tabular}{llllll}\hline
& \multicolumn{2}{c}{$\frac{\langle
{\bar K}^0 \vert \hat O_{VV+AA} \vert K^0 \rangle }
{\vert \langle 0 \vert {\hat P} \vert K^0 \rangle \vert^2}$ 
at $m_q=0$} & \multicolumn{2}{c}{$B_K($NDR, 2GeV)} \\
    $\beta$       & WI  & PT  & WI  & WI[VS]  & PT \\ 
\hline
5.9               &$-0.0200(39)$  &$-0.0415(8)$  
                  &$+0.38(6)$     &$+0.168(20)$   &$-0.468(14)$ \\ 
6.1               &$-0.0068(55)$  &$-0.0333(10)$  
                  &$+0.68(11)$    &$+0.288(29)$   &$-0.225(22)$ \\ 
6.3               &$-0.0017(74)$  &$-0.0240(12)$  
                  &$+0.69(12)$    &$+0.342(33)$   &$-0.000(21)$ \\ 
6.5               &$+0.006(10)$   &$-0.0188(17)$  
                  &$+0.72(18)$    &$+0.360(52)$   &$+0.156(40)$ \\ 
\hline
$a=0$             &               &$-0.0009(31)$  
                  &               &$+0.562(64)$   &$+0.639(76)$ \\ 
\hline
\end{tabular} 
\end{center}
\end{table}

\newpage
\begin{center}
\section*{Figure Captions}
\end{center}
\begin{itemize}
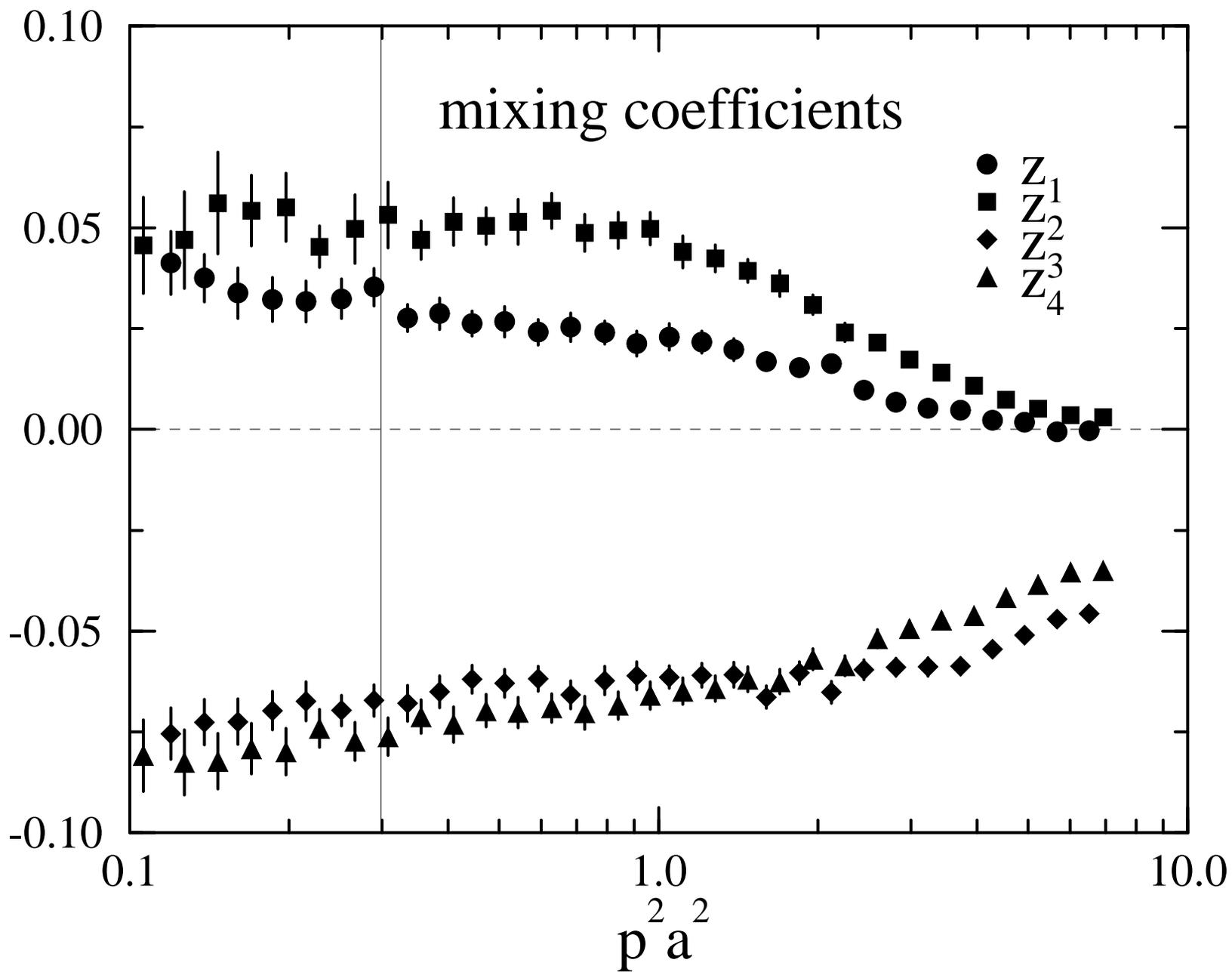
\item[Fig.~1]  Mixing coefficients $z_1,\cdots,z_4$ plotted as 
a function of external momentum squared $(pa)^2$   
for $K=0.15034$ at $\beta=6.3$. Vertical line corresponds to 
$p^*\approx 2$ GeV.

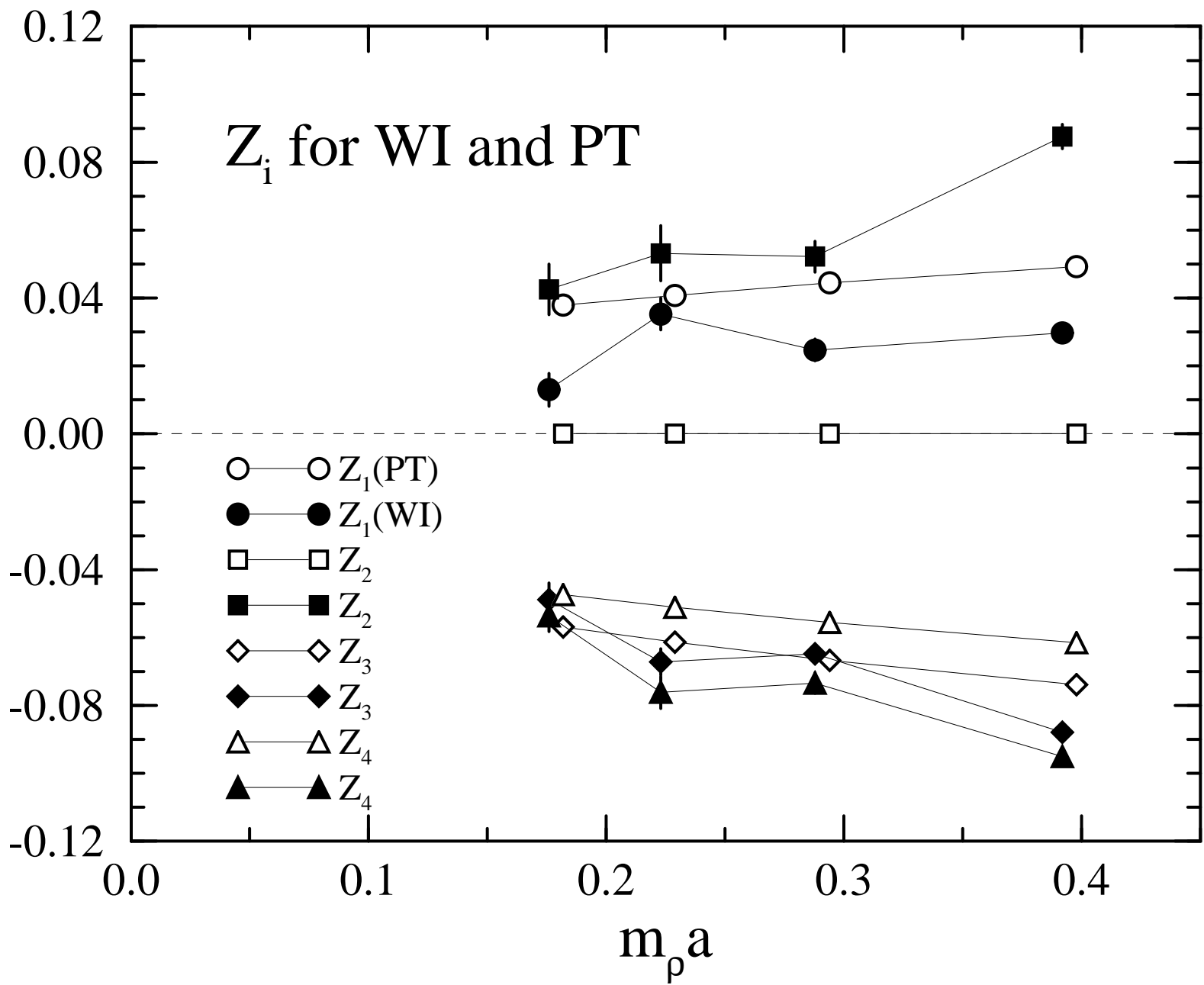
\item[Fig.~2] Comparison of the mixing coefficients
$z_1,\cdots,z_4$ evaluated at ${p^{*}}\approx 2$ GeV 
using the  Ward identity (WI; solid symbols) and perturbative (PT;
open symbols) methods. The coefficients are plotted 
as a function of $m_\rho a$. 

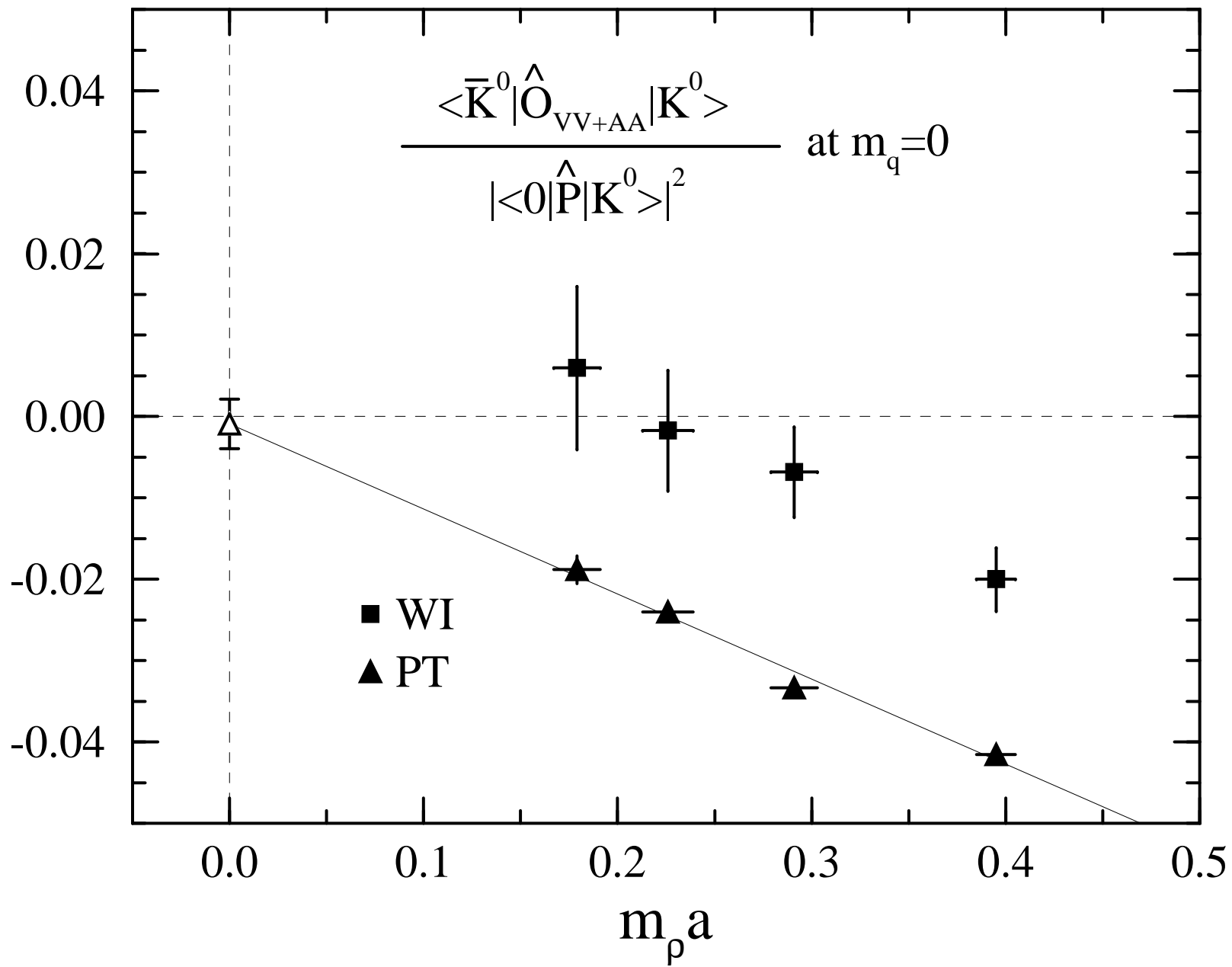
\item[Fig.~3] Test of the chiral behavior of 
$\langle\bar K^0|\hat O_{VV+AA}|K^0\rangle/
|\langle 0|\hat P|K^0\rangle|^2$ at $m_q=0$
for the WI and PT methods. 
The operators are renormalized at 2 GeV in the NDR scheme.
For both methods we use the same $\hat P$ perturbatively corrected
with the tadpole improvement.
The solid line is a linear extrapolation to the continuum 
limit.

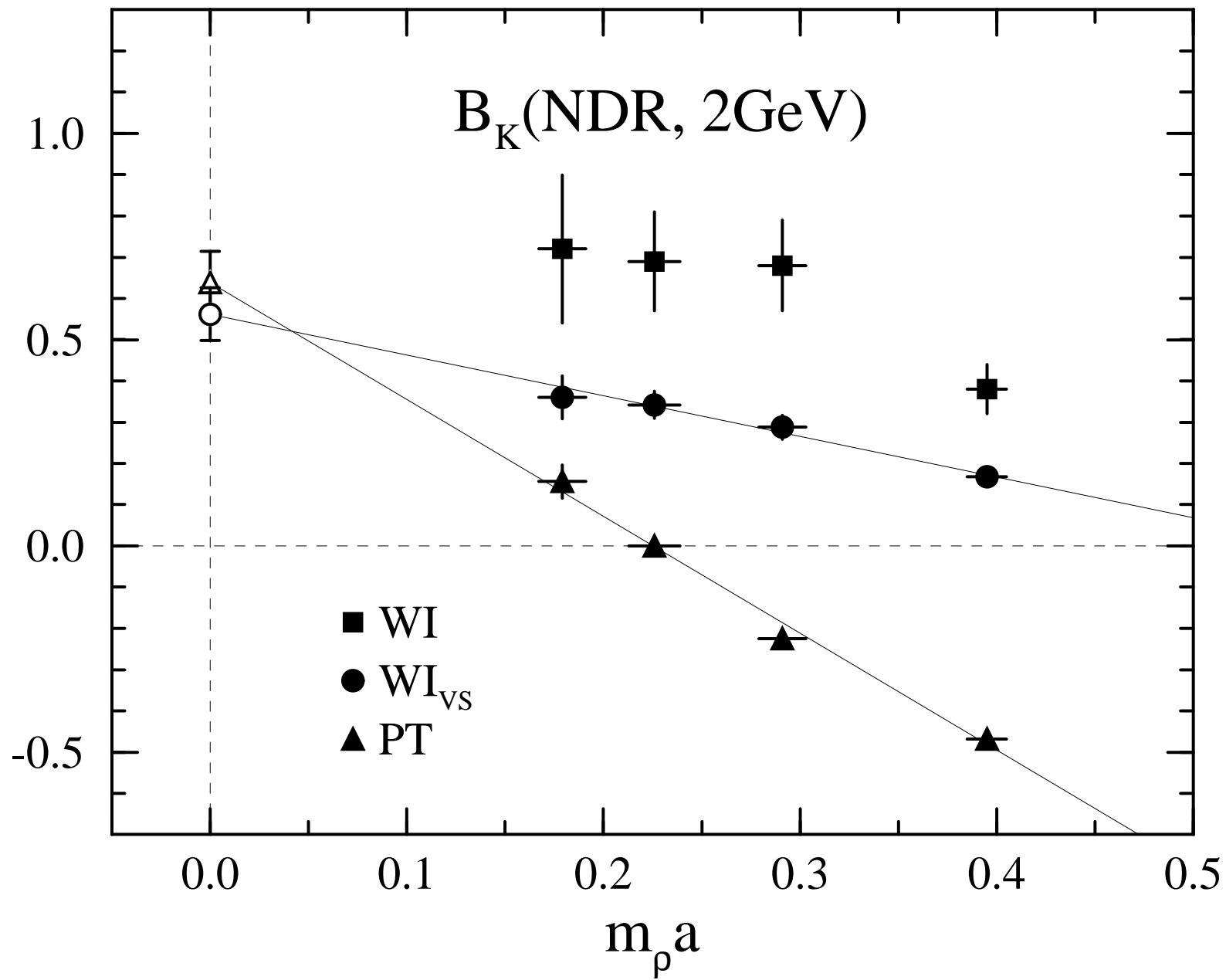
\item[Fig.~4] $K^0-\bar K^0$ matrix element  
$B_K($NDR, 2GeV) plotted as a function of the lattice
spacing for the WI, WI$_{\rm VS}$ and PT methods. 
The solid lines show linear 
extrapolations to the continuum limit. 

\end{itemize}



 
\end{document}